\documentclass[conference]{IEEEtran}

\usepackage{soul}

\usepackage[T1]{fontenc}
\usepackage{bm}
\usepackage{amssymb,amsmath,amsfonts,amsthm}
\usepackage{mathrsfs}
\usepackage{cite}
\usepackage{array}
\usepackage{tabularx}
\usepackage[table]{xcolor}
\usepackage{multicol, multirow}
\usepackage{color}
\usepackage{mathtools}
\usepackage{subcaption}
\usepackage{tikz,pgf}
\usepackage{verbatim}
\usetikzlibrary{calc,positioning,mindmap,trees,decorations.pathreplacing}
\usepackage{graphicx}
\usepackage{bbm}
\usepackage[utf8]{inputenc}
\usepackage{algorithm}
\usepackage{algorithmic}
\usepackage{hyperref}
\usepackage{pgfplots}
\pgfplotsset{compat=newest}
\usepackage[acronym,shortcuts]{glossaries-extra}
\glsdisablehyper
\setabbreviationstyle[acronym]{long-short}

\DeclarePairedDelimiter\abs{\lvert}{\rvert}
\DeclarePairedDelimiter\norm{\lVert}{\rVert}

\usepackage[colorinlistoftodos,prependcaption,backgroundcolor=black!5!white,bordercolor=red]{todonotes}

\DeclareMathOperator{\Tr}{Tr}
\newacronym{isac}{ISAC}{integrated sensing and communication}
\newacronym{sinr}{SINR}{signal-to-interference-noise ratio}
\newacronym{csi}{CSI}{channel state information}
\newacronym{uav}{UAV}{unmanned aerial vehicle}
\newacronym{ssb}{SSB}{synchronization signal block}
\newacronym{mimo}{MIMO}{multiple-input multiple-output}

\makeatletter
\let\oldabs\abs
\def\abs{\@ifstar{\oldabs}{\oldabs*}}

\IEEEoverridecommandlockouts
\ifCLASSINFOpdf
\else
\fi

\usepackage[cmintegrals]{newtxmath}
\def\BibTeX{{\rm B\kern-.05em{\sc i\kern-.025em b}\kern-.08em
    T\kern-.1667em\lower.7ex\hbox{E}\kern-.125emX}}
\makeglossaries

\begin{document}

\title{Drone Surveillance via Coordinated Beam Sweeping in MIMO-ISAC Networks} %

\author{\IEEEauthorblockN{ Palatip Jopanya, Diana P. M. Osorio, and Erik G. Larsson} \IEEEauthorblockA{ {Department of Electrical Engineering,} {Link\"oping University}, Sweden }  \IEEEauthorblockA{E-mail: \{palatip.jopanya, diana.moya.osorio, erik.g.larsson\}@liu.se}

}

\maketitle

\begin{abstract}
This paper introduces a scheme for drone surveillance coordinated with the fifth generation (5G) \gls{ssb} cell-search procedure to simultaneously detect low-altitude drones within a volumetric surveillance grid. Herein, we consider a multistatic configuration where multiple access points (APs) collaboratively illuminate the volume while independently transmitting \Gls{ssb} broadcast signals. Both tasks are performed through a beam sweeping. In the proposed scheme, coordinated APs send sensing beams toward a grid of voxels within the volumetric surveillance region simultaneously with the 5G \Gls{ssb} burst. To prevent interference between communication and sensing signals, we propose a precoder design that guarantees orthogonality of the sensing beam and the \Gls{ssb} in order to maximize the sensing signal-to-interference-plus-noise ratio (SINR) while ensuring a specified SINR for users, as well as minimizing the impact of the direct link. The results demonstrate that the proposed precoder outperforms the non-coordinated precoder and is minimally affected by variations in drone altitude.
\end{abstract}

\section{Introduction}
With the rise of \gls{isac} systems, drone detection has emerged as a promising capability, as highlighted in the feasibility study report by the 3rd Generation Partnership Project (3GPP)~\cite{TR}. However, reliable detecting small aerial targets, such as drones, with \Gls{isac} system remains a challenge.

To explore the potential for enabling this capability, it is essential to investigate which signals can be effectively utilized for sensing. In current \gls{mimo} cellular systems, communication signals are of two types: user-specific signals that use transmitter \gls{csi} and signals that do not use transmitter \Gls{csi}, e.g. broadcast signal type such as synchronization signal block (\gls{ssb}). Many studies, such as~\cite{detunauth,10742291}, assume the availability of transmitter-side \gls{csi} for precoder design in order to ensure that the sensing signal does not interfere with the signal intended for communication users.  However, there are limited studies targeting the use of broadcast reference signals in \Gls{isac} systems. 

In \cite{10200933}, the authors propose an ISAC signal model that uses reference signals, specifically the \gls{ssb} and SIB1, as illuminators for sensing. The range-velocity profile yields a higher peak-to-sidelobe ratio by including more symbols (SIB1 symbols) for illumination. In \cite{10083170}, the authors present a passive coherent location system that uses \Gls{ssb} and traffic symbols to enable more symbols for illumination. In \cite{utilizingSSB}, the authors repurposed \Gls{ssb} signals by adding additional \Gls{ssb} beams, which are used for drone detection and localization. The common approach in these works is the reuse of reference-type communication signals without the addition of dedicated sensing signals. These approaches are effective for targets within the base station's coverage, but may not perform well in detecting low-altitude drones, which are small and often positioned in weak coverage areas, as the system is primarily designed for connectivity of user equipments (UEs).

In \cite{blandino}, the authors investigate unmanned aerial vehicle (UAV) detection and positioning using standardized positioning reference signal (PRS) signals in a monostatic setting. The results show that detection performance depends on the drone altitudes in both urban macro and urban micro scenarios.

\emph{Contributions:} To address these gaps, we propose to include a dedicated sensing signal, which is coordinated with the \Gls{ssb} signals of 5G NR,
in  MIMO cellular systems. In this scheme, at each symbol in the \Gls{ssb}, a sensing signal is added and directed toward a voxel in a predetermined three-dimensional volume for surveillance. This task is performed through the collaboration of multiple coordinated access points (APs), in a multistatic setting, providing multiple illuminators and a single sensing receiver. To this end, a sensing precoder is designed to maximize the sensing SINR, while ensuring orthogonality with both the SSB precoder and the direct communication link. This design  maintains the user’s SINR, preserving communication quality.\footnote{For reproducibility of this work, the code is available in \cite{code}.}

\textbf{Notation:} $(\cdot)^T$,$(\cdot)^\ast$, and $(\cdot)^H$ denote transpose, conjugate, and Hermitian transpose, respectively. Lowercase boldface characters are column vectors and uppercase boldface characters are matrices. $\Tr\{\cdot\}$ denotes the trace of a matrix. $[\cdot]_i$ is the $i$th element of a vector, $\mathbb{E\{ \cdot \}}$ is the expected value, $\norm{\cdot}$ is the Euclidean norm, and $\otimes$ is the Kronecker product.

\section{System model}
\begin{figure}[htbp]
    \centering
        \includegraphics[width=3.5in]{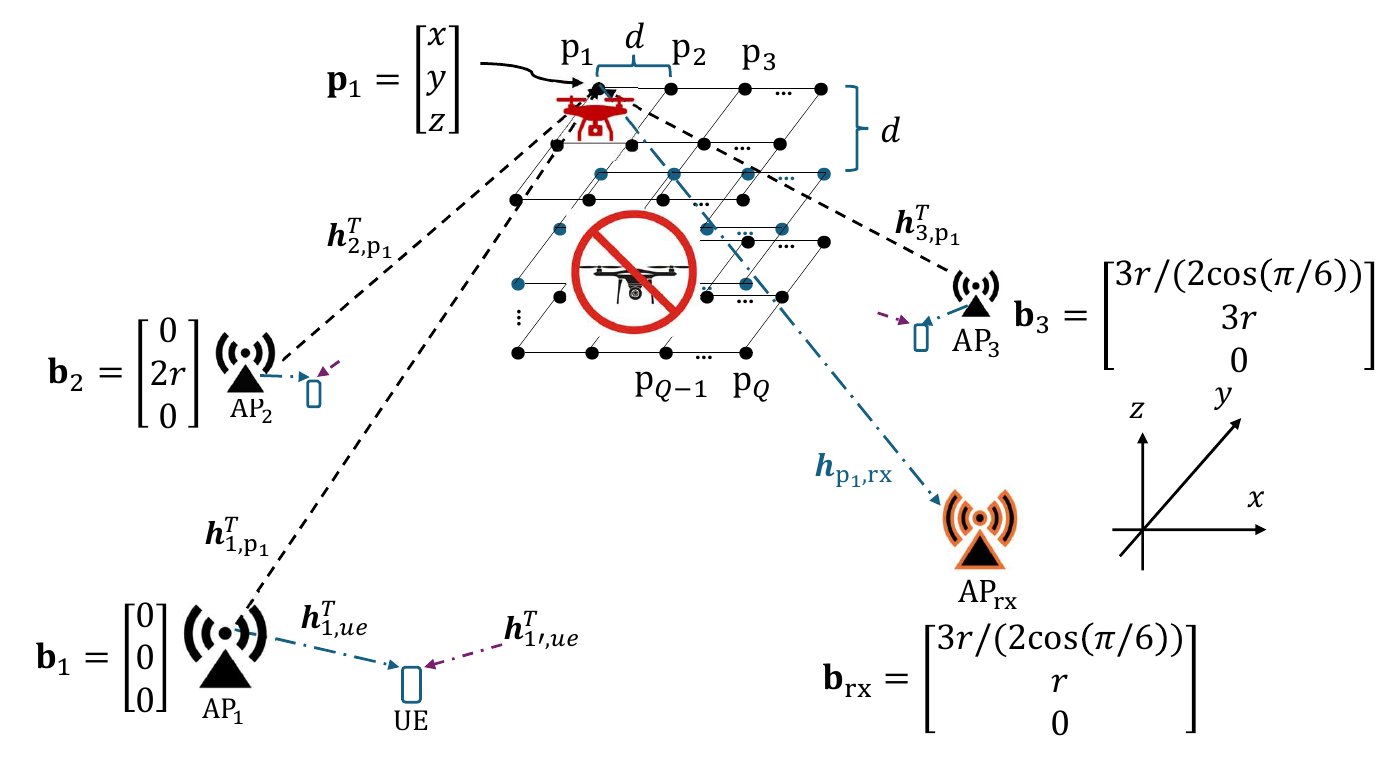}
        \caption{Multistatic setup in a hexagonal grid.}
    \label{fig:system_model}
\end{figure}

Consider the \gls{mimo}-\gls{isac} network illustrated in Figure \ref{fig:system_model}, which depicts $J$ \Gls{isac} APs in a cellular network, which are arranged in a hexagonal grid with inner radius $r$. Besides providing connectivity to users within their cells, the APs collaborate for performing sensing of low-altitude drones. This task is achieved by having several illuminator $(\text{AP}_1,\dots,\text{AP}_J)$ and one sensing receiver, $\text{AP}_\text{rx}$, in a multistatic sensing setting. The surveillance region is a collection of $Q$ equidistant voxels with spacing $d$ forming a three-dimensional volume. Additionally, each AP is equipped with a uniform planar array (UPA) with $M $$=$$ M_\text{V} \times M_\text{H}$ antennas, with $M_\text{H}$ antennas per row and $M_\text{V}$ antennas per column and $M_\text{V} $$=$$ M_\text{H}$. The boresight direction is along the $x$-axis, toward the positive $x$-axis for panel $\text{AP}_1$ and $\text{AP}_2$, and toward the negative $x$-axis for panel $\text{AP}_3$ and $\text{AP}_{\text{rx}}$. 
$\text{AP}_j$ is located at the three-dimensional Cartesian coordinate $\textbf{b}_j \in \mathbb{R}^3$; that is $[\textbf{b}_j]_1$ in $x$,  $[\textbf{b}_j]_2$ in $y$ and $[\textbf{b}_j]_3$ in $z$. Similarly, $\text{AP}_\text{rx}$ is located at $[\textbf{b}_\text{rx}]_1$ in $x$,  $[\textbf{b}_\text{rx}]_2$ in $y$ and $[\textbf{b}_\text{rx}]_3$ in $z$.

\begin{figure}[ht]
        \centering
            \includegraphics[width=3.5in]{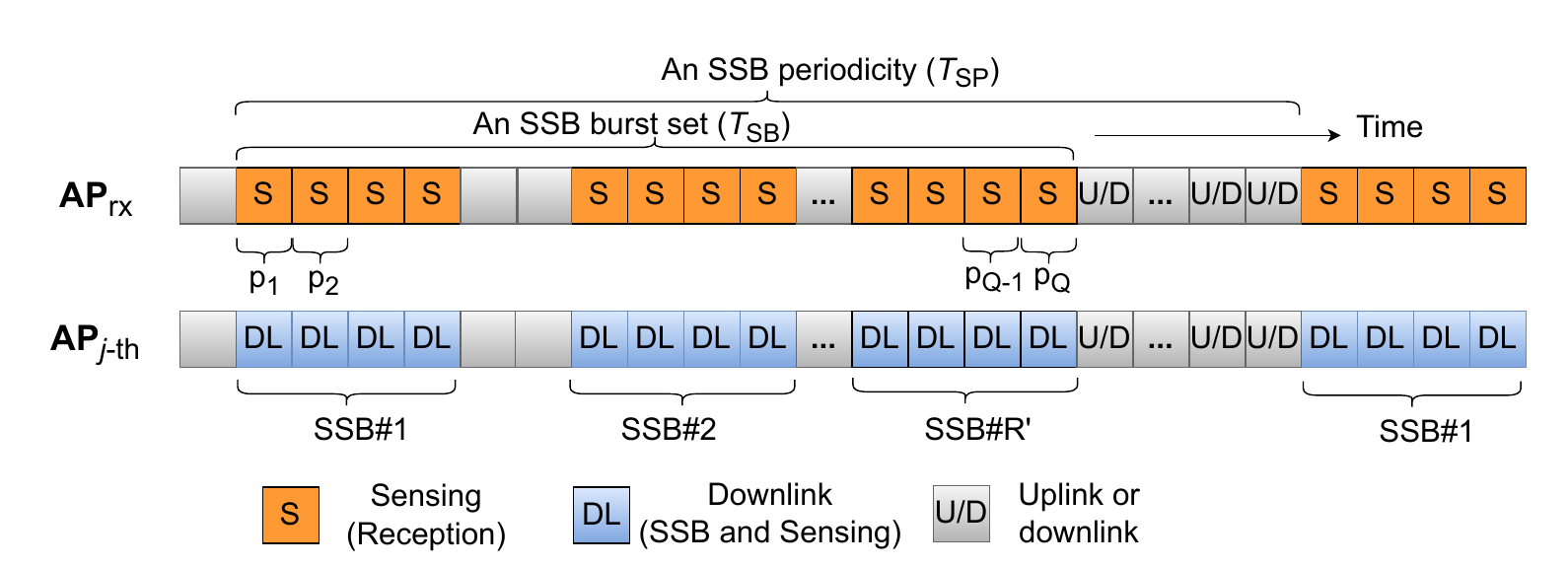}
            \caption{TDD flow with sensing reception with $R$$=$$Q$.}
        \label{fig:tdd_flow}
\end{figure} 
\subsection{Proposed ISAC scheme}
In the proposed scheme, a sensing signal is transmitted through an orthogonal sensing beam concurrently with the SSB symbol at each  $\text{AP}_j$ and in a coordinated manner among different APs. These sensing beams sweep across voxels within the designated volume at altitudes where low-altitude drones may fly, as illustrated in Fig.~\ref{fig:system_model}. However, this setup introduces the possibility of interference between the sensing signal and the SSB signal at the user. To mitigate this, a precoder is designed to suppress such interference, as detailed in the following section. To enable this, the system should implement an adaptation of the time-division duplexing (TDD) workflow, where $\text{AP}_\text{rx}$ switches to reception mode when $\text{AP}_j$ transmits \Gls{ssb} and sensing symbols, as illustrated in Fig. \ref{fig:tdd_flow}. Each $\text{AP}_j$ performs beam sweeping of \Gls{ssb} signals with time multiplexing toward its own UE's coverage in a cellular-based system. 

\subsection{SSB signals}
An \Gls{ssb} is a reference-type signal that enables user equipment to synchronize with the network for initial access and mobility. This is structured as a block of OFDM symbols \cite{5gnr}. Each block (beam) is transmitted in a distinct directional beam, and each individual \Gls{ssb} is separated from the next by a time gap to allow the antenna system to reconfigure its beam direction and to avoid time-domain interference. A set of \Gls{ssb} blocks, called an \Gls{ssb} burst set, covers the entire AP's coverage. This is done in a beam sweeping process, which is confined within $T_{\text{SB}}$ (\Gls{ssb} burst set period) and will be repeated periodically in every \Gls{ssb} periodicity, $T_{\text{SP}}$. 
We denote the normalized steering matrix for \Gls{ssb} signals, $\mathbf{F} $$\in$$ \mathbb{C}^{M \times R}$, as
\begin{equation}
\begin{split}
\label{eqn:fmat}
\mathbf{F} =& \frac{1}{\sqrt{M}} [ \mathbf{a} (\theta_1,\phi_1),\mathbf{a} (\theta_1,\phi_1),\mathbf{a} (\theta_1,\phi_1),\mathbf{a} (\theta_1,\phi_1),\cdots \\
&\mathbf{a} (\theta_{R'},\phi_{R'}),\mathbf{a} (\theta_{R'},\phi_{R'}),\mathbf{a} (\theta_{R'},\phi_{R'}),\mathbf{a} (\theta_{R'},\phi_{R'}) ] ,
\end{split}
\end{equation}
\noindent
where $R'$ is the total number of \Gls{ssb} blocks in a burst set, $R$$=$$4R'$ is the total number of symbols of all \Gls{ssb} blocks in a burst set (each block has a width of 4 symbols). The half-wavelength-spaced antenna response of the UPA is $\mathbf{a} (\theta,\phi) $$=$$ \mathbf{a}_{M_\text{V}}(\phi,0)\otimes \mathbf{a}_{M_\text{H}}(\theta,\phi)$, where $\theta$ is the azimuth, $\phi$ is the elevation, $\mathbf{a}_{M_\text{V}} (\phi,0) $$=$$ [1, e^{-j\pi \sin(\phi)},\cdots, e^{-j \pi  ({M}_V-1)\sin(\phi)}]^T$, $\mathbf{a}_{M_\text{H}} (\theta,\phi) $$=$$ [1, e^{-j \pi \sin(\theta)\cos(\phi)},\cdots, e^{-j \pi ({M}_H-1)\sin(\theta)\cos(\phi)}]^T$, $\mathbf{a}_{M_\text{H}} (\theta,\phi) $$\in$$ \mathbb{C}^{M_H}$, $\mathbf{a}_{M_\text{V}} (\phi,0) $$\in$$ \mathbb{C}^{M_V}$ and $\mathbf{a} (\theta,\phi) $$\in$$ \mathbb{C}^{M}$ \cite{IntroIRS}. We obtain a set of angle pairs, $\{(\theta,\phi)\}$, for the columns in $\mathbf{F}$ with $\theta =  \{\arcsin{( {2i}/{\sqrt{M}} )} \}$ and $\phi = \{ \arcsin{( {2i'}/{\sqrt{M}} )} \}$ for $i $$\in$$ \{ 0,\pm1,\pm2,.., \pm\lfloor {\sqrt{M}}/{2}  \rfloor \}$, and $i' $$\in$$ \{ 0,-1,-2,.., -\lfloor {\sqrt{M}}/{2}  \rfloor \}$.

\subsection{Sensing Channel}
\begin{figure}[ht]
        \centering
        \includegraphics[width=2.8in]{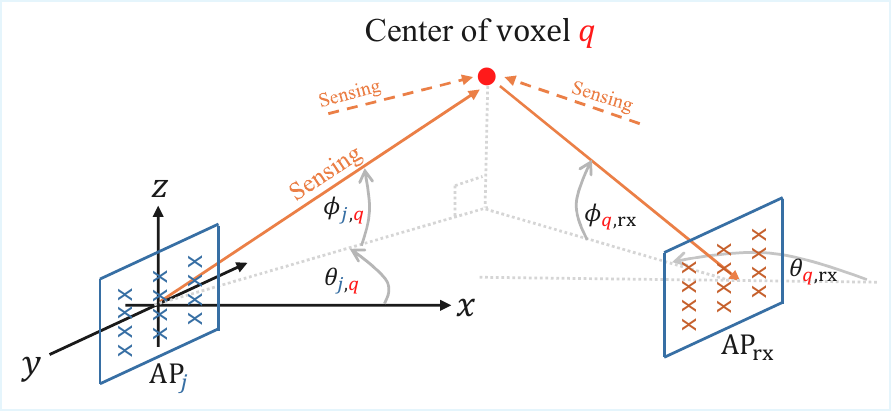}
            \caption{Departure and arrival angles for the sensing channel of voxel index $q$.}
        \label{fig:angles}
\end{figure}

The two-link channel at voxel $q$, $\textbf{H}_{j,q}\in \mathbb{C}^{M\times M}$ is the far-field LOS channel between $\text{AP}_j$-$q$-$\text{AP}_\text{rx}$ and is defined as 
\begin{equation}
    \label{eqn:deter_cha}
    \textbf{H}_{j,q}= \eta \sqrt{\beta_{j,q}} e^{-j2\pi f_c \tau_{j,q} } \Tilde{\textbf{H}}_{j,q} ,
\end{equation}
where $\eta $$\in$$ \{0,1\}$; $\eta$$=$$0$ if the drone is not present at the voxel $q$, $\eta$$=$$1$ if the drone is present at the voxel, $f_c$ is the carrier frequency, $\tau_{j,q}$ is the propagation delay of the bistatic link, $\beta_{j,q} $$=$$ {1}/{(l^2_{j,q}l^2_{q,\text{rx}} )}$ is the normalized large-scale fading of the bistatic link \cite[Chap. 2]{doi:10.1049/SBRA021E}, $l_{j,q}$ is the distance between $\text{AP}_j$ and voxel $q$, and $l_{q,\text{rx}}$ is the distance between $q$ and $\text{AP}_\text{rx}$. 
The LOS channel matrix is $\Tilde{\textbf{H}}_{j,q} $$=$$ \textbf{h}_{q,\text{rx}}\textbf{h}_{j,q}^T$, where $\textbf{h}_{q,\text{rx}}$ is the channel from voxel $q$ to $\text{AP}_\text{rx}$ and is defined as $\textbf{h}_{q,\text{rx}} $$=$$ \mathbf{a}
(\theta_{q,\text{rx}},\phi_{q,\text{rx}})$, with 
\begin{equation}
    \begin{split}
    \theta_{q,\text{rx}} &=\arctan2({[\textbf{b}_\text{rx}]_2-[\textbf{p}_q]_2},{[\textbf{b}_\text{rx}]_1-[\textbf{p}_q]_1} \Bigl),\\
    \phi_{q,\text{rx}} &= \arcsin ( ( [\textbf{b}_\text{rx}]_3-[\textbf{p}_q]_3 )/l_{q,\text{rx}} ),
    \end{split}
\end{equation}
for $-\pi/2 \leq \phi_{q,\text{rx}} \leq \pi/2$, $l_{q,\text{rx}}$$=$$\norm{\textbf{p}_q-\textbf{b}_\text{rx}}$, and $\textbf{p}_q$ is the three-dimensional Cartesian coordinate of the center of the voxel $q$ $\in \mathbb{R}^3$. Similarly, $\textbf{h}_{j,q}$ is the channel from $\text{AP}_j$ to voxel $q$ and is defined as $\textbf{h}_{j,q} $$=$$ \mathbf{a} (\theta_{j,q},\phi_{j,q})$, with
\begin{equation}
    \begin{split}
        \theta_{j,q} &= \arctan2({[\textbf{p}_q]_2-[\textbf{b}_j]_2},{[\textbf{p}_q]_1-[\textbf{b}_j]_1} ), \\
        \phi_{j,q} &= \arcsin ( {( [\textbf{p}_q]_3-[\textbf{b}_j]_3 )}/{l_{j,q}} ),
    \end{split}
\end{equation}
for $-\pi/2 \leq \phi_{j,q} \leq \pi/2$ and $l_{j,q} = \norm{\textbf{p}_q-\textbf{b}_j}$. The angles for sensing channel are illustrated in Fig. \ref{fig:angles}.

\subsection{Direct-link channel}
The direct-link channel, $\mathbf{H}_{0,j}$, is the LOS channel between $\text{AP}_j$ and $\text{AP}_\text{rx}$ and is modeled as 
\begin{equation}
    \label{eqn:direct_chan}
    \textbf{H}_{0,j}= \sqrt{\beta_{0,j}} e^{-j2\pi f_c \tau_{0,j} } \Tilde{\textbf{H}}_{0,j},
\end{equation}
where $\tau_{0,j}$ is the propagation delay of the direct-link, $\beta_{0,j}={1}/{l _{0,j}^2}$ is the normalized large-scale fading of the direct-link, and $l_{0,j}$ is the distance between $\text{AP}_j$ and $\text{AP}_\text{rx}$. The LOS channel matrix is $\Tilde{\textbf{H}}_{0,j} $$=$$ \textbf{h}_{0,j,\text{a}}\textbf{h}_{0,j,\text{d}}^T$, where $\textbf{h}_{0,j,\text{a}}$$=$$\textbf{a}(\theta_{\text{a},j},0)$, $\textbf{h}_{0,j,\text{d}}$$=$$\textbf{a}(\theta_{\text{d},j},0)$, $\theta_{\text{d},j}$ is the departure azimuth from $\text{AP}_j$, $\theta_{\text{a},j}$ is the arrival azimuth at $\text{AP}_\text{rx}$ from $\text{AP}_j$, and $\theta_{\text{a},j}=\theta_{\text{d},j}+\pi$.
\subsection{Transmitted signal}
We consider the case where $Q$$<$$R$ and we index the symbols using the voxel indices $q$. The transmitted signal, $\mathbf{x}_{j}[q] $$\in $$\mathbb{C}^M$, of a single subcarrier at $\text{AP}_j$ and at the $q$th symbol is 
\begin{equation}
\label{eqn:tx_apj}
\mathbf{x}_{j}[q] = \underbrace{ s[q] \mathbf{w}_{j,q} }_\text{Sensing signal} + \underbrace{\sqrt{\rho_{j,q}} c_{j} [q] \mathbf{f}^\ast_{q} }_\text{\Gls{ssb} signal},
\end{equation}
where $\mathbf{w}_{j,q} $$\in$$ \mathbb{C}^M$ is the sensing precoder toward voxel $q$, $\mathbf{f}^{\ast}_q $$\in$$ \mathbb{C}^M$ is the \Gls{ssb} precoder, $s[q]$$=$$s_{q} e^{j2\pi f_cqT}$ is the sampled sensing signal, $T$ is the symbol duration, $c_{j}[q] $$=$$ c_{q,j} e^{j2\pi f_cqT}$ is the sampled \Gls{ssb} signal, $\mathbf{f}_{q} $$\in$$ \mathbb{C}^M$ is $q$th column in (\ref{eqn:fmat}), $s_{q}$ is an uncorrelated complex sensing symbol with unit power shared by all $\text{APs}$, $c_{q,j}$ is an uncorrelated complex \Gls{ssb} symbol with unit power, and  $\rho_{j,q}$ is the signal power of the \Gls{ssb} at symbol $q$th. For simplicity, let $q=\{1,2,..,Q\}$ denote the symbol indices, considering only the slots that contain \Gls{ssb} symbols and disregarding the symbols between each block.

\subsection{Received signal}
The received signal, $\mathbf{y}_{\text{ap}}[q] $$\in$$ \mathbb{C}^{M}$, at $\text{AP}_\text{rx}$ and the $q$th symbol is given by
\begin{equation}
\begin{split}
    \mathbf{y}_{\text{ap}}[q] &= \sum_{j=1}^{J} \alpha_{j} \textbf{H}_{j,q}\mathbf{x}_{j}[q] + \sum_{j=1}^{J} \textbf{H}_{0,j}\mathbf{x}_{j}[q] \\
    &+ \sum_{j=1}^{J}  \textbf{G}_{j,q}\mathbf{x}_{j}[q] + \mathbf{n}[q],
    \label{eqn:y_apr_all}
\end{split}
\end{equation}
where the first term accounts for echoes from the voxel index $q$ with RCS that follows Swerling 2 model (\emph{SW2}), $\alpha_{j} $$\sim$$ \mathcal{CN}(0,\sigma_{\text{rcs}}^2)$, the second term accounts for direct-link, the third term accounts for the ground clutter with $\mathbf{G}_{j,q} $$\sim$$ \mathcal{CN}(0,\beta_{\text{g},q}\mathbf{I})$ being the scattering channel of the ground clutter, and $\mathbf{n}[q] $$\sim$$ \mathcal{CN}(0,\sigma^2_\text{n}\mathbf{I})$ is additive white Gaussian noise (AWGN). Equation ($\ref{eqn:y_apr_all}$) can be re-expressed by considering the direct link, noise, and the dominant echoes from voxel and ground clutter (SSB echoes) as 
\begin{equation}
\begin{split}
\label{eqn:rx_apr}
\mathbf{y}_{\text{ap}}[q] &= \underbrace{s[q]\sum_{j=1}^{J} \alpha_{j} \textbf{H}_{j,q} \mathbf{w}_{j,q}}_\text{Sensing echoes} + \underbrace{\sum_{j=1}^{J}
\sqrt{\rho_{j,q}} c_{j} [q] \textbf{G}_{j,q}\mathbf{f}^\ast_{q} }_\text{\Gls{ssb} echoes (clutter)} \\ 
&+ \underbrace{\sum_{j=1}^{J}\mathbf{H}_{0,j} (s[q]\mathbf{w}_{j,q} + \sqrt{\rho_{j,q}} c_{j} [q]\mathbf{f}^\ast_{q}) }_\text{Direct-link}  + \underbrace{\mathbf{n}[q]}_\text{Noise},\\
\end{split}
\end{equation}
where the total transmitted power at $\text{AP}_j$ is $\norm{\mathbf{w}_{j,q}}^{2} $$+$$ \rho_{j,q}$. Define a combining vector, $\mathbf{v}_q $$\in$$ \mathbb{C}^M$, as $\mathbf{v}_q = \mathbf{h}_{q,\text{rx}}/\sqrt{M}$. The received signal in (\ref{eqn:rx_apr}) with the combining vector, $y'_{\text{ap}}[q]$, is $y'_{\text{ap}}[q] = \mathbf{v}^H_q\mathbf{y}_{\text{ap}}[q]$.

At UE side, the received signal, $y_{\text{ue},j}[q] $$\in$$ \mathbb{C}$, of $q$th symbol at UE of $\text{AP}_j$ is
\begin{equation}
    \begin{split}
    \label{eqn:y_ue}
    y_{\text{ue},j}[q] &= \underbrace{\sqrt{\rho_{j,q}} c_{j} [q] \mathbf{h}_{j,\text{ue}}^T\mathbf{f}^\ast_{q} }_\text{Desired signal (\Gls{ssb})} + \underbrace{\sum_{i\in\mathcal{U}}\sqrt{\rho_{i,q}} c_{i} [q]\mathbf{h}_{i,\text{ue}}^{T}\mathbf{f}^\ast_{q} }_\text{Interference (\Gls{ssb})} \\
    &+ \underbrace{ s[q] \sum_{k=1}^{J}   \mathbf{h}_{k,\text{ue}}^T\mathbf{w}_{k,q} }_\text{Interference (Sensing)} + \underbrace{n[q]}_\text{Noise},
    \end{split}
\end{equation}
where $\mathbf{h}_{j,\text{ue}} $$\sim$$ \mathcal{CN}(0,\beta_{j,\text{ue}}\mathbf{I})$, $\mathcal{U}$ is is the set of interfering APs, and $n[q] $$\sim$$ \mathcal{CN}(0,\sigma^2_\text{n})$ is   noise. Given that ${u}[q]$ is an undesired part of the received signal at the UE (from the second to the fourth terms in (\ref{eqn:y_ue})), the received signal, $y_{\text{ue},j}[q]$, can be rewritten as $y_{\text{ue},j}[q] = \sqrt{\rho_{j,q}}c_{j} [q]\mathbf{h}_{j,\text{ue}}^T\mathbf{f}^\ast_{q}  + {u}[q]$.
\subsection{Average sensing and user SINR}
Two sensing performance metrics are considered: average sensing SINR with and without direct link, which are used in simulation and optimization, respectively. The average sensing SINR, $\gamma_{\text{sen},q}$, at $\text{AP}_\text{rx}$ at $q$th symbol considering direct link is 
\begin{equation}
    \label{eqn:SNR_sensing}
    \gamma_{\text{sen},q} = \frac{\mathbb{E} \Bigl\{ \abs{  s[q] \sum_{j=1}^{J} \alpha_{j} \mathbf{v}^H_q\textbf{H}_{j,q} \mathbf{w}_{j,q}}^{2} \Bigl\} }{\mathbb{E} \Bigl\{ \abs{ \mathbf{v}^H_q( \mathbf{d}[q] + \sum_{j=1}^{J}\sqrt{\rho_{j,q}}c_{j} [q]\textbf{G}_{j,q}\mathbf{f}^\ast_{q}  + \mathbf{n}[q])}^{2} \Bigl\} },
\end{equation}
where $\mathbf{d}[q]=\sum_{j=1}^{J}\mathbf{H}_{0,j} (s [q]\mathbf{w}_{j,q} + \sqrt{\rho_{j,q}} c_{j} [q]\mathbf{f}^\ast_{q})$ is the direct link term in (\ref{eqn:rx_apr}), and $\eta $$=$$ 1$. 
The average $\gamma_{\text{sen},q}$ in (\ref{eqn:SNR_sensing}), disregarding the direct link, is
\begin{equation}
    \begin{split}
    \label{eqn:SCNR_sensing}
    \gamma_{\text{sen},q}' &= \frac{\mathbb{E} \Bigl\{ \abs{  s[q] \sum_{j=1}^{J} \alpha_{j} \mathbf{v}^H_q\textbf{H}_{j,q} \mathbf{w}_{j,q}}^{2} \Bigl\} }{\mathbb{E} \Bigl\{ \abs{ \mathbf{v}^H_q(  \sum_{j=1}^{J}\sqrt{\rho_{j,q}}c_{j} [q]\textbf{G}_{j,q}\mathbf{f}^\ast_{q}  + \mathbf{n}[q])}^{2} \Bigl\} } \\
    &=  \frac{\sigma^2_{\text{rcs}}\abs{\mathbf{v}^H_q \textbf{H}_{q} \mathbf{w}_{q}}^2} { \beta_{\text{g},q}\sum_{j=1}^{J}\rho_{j,q} + \sigma_n^2},
    \end{split}
\end{equation}
where $\textbf{H}_{q}$$\in$$ \mathbb{C}^{M\times JM}$ is $\textbf{H}_{q}$$=$$[\textbf{H}_{1,q},\cdots,\textbf{H}_{J,q}]$ and $\mathbf{w}_{q} $$\in$$ \mathbb{C}^{JM}$ is $\mathbf{w}_{q}$$=$$[\mathbf{w}^T_{1,q},\cdots,\mathbf{w}^T_{J,q}]^T$. Note that $\textbf{v}_q^H$ and $\textbf{f}_q$ are unit-norm vectors.

The average SINR of \Gls{ssb} signal at UE in $\text{AP}_j$'s coverage at $q$th symbol is 
\begin{equation}
\begin{split}
    \label{eqn:SINR_UE}
    \gamma_{\text{ue},j,q} = \frac{\mathbb{E} \Bigl\{ \abs{\sqrt{\rho_{j,q}} c_{j} [q]\mathbf{h}_{j,\text{ue}}^T\mathbf{f}^\ast_{q} }^{2} \Bigl\} }{\mathbb{E} \Bigl\{ \abs{{u}[q] }^{2} \Bigl\} } \\
    = \frac{ \rho_{j,q} \beta_{j,\text{ue}}}{ \sum_{i\in \mathcal{U}} \rho_{i,q}  \beta_{i,\text{ue}} + \sum_{k=1}^{J}\beta_{k,\text{ue}}\norm{\mathbf{w}_{k,q}}^2+\sigma_n^2}.
\end{split}
\end{equation}
For simplicity, the sensing power at the UE, $\beta_{k,\text{ue}}\norm{\mathbf{w}_{k,q}}^2$, is neglected in this paper, and $(\ref{eqn:SINR_UE})$ is rewritten as  
\begin{equation}
    \label{eqn:SINR_UE_3}
    \gamma_{\text{ue},j,q} = \frac{\rho_{j,q} \beta_{j,\text{ue}}}{ \sum_{i\in \mathcal{U}} \rho_{i,q}  \beta_{i,\text{ue}} + \sigma_n^2}.
\end{equation}
Note that the average $\gamma_{\text{ue},j,q}$ is identical for all $\text{AP}_j$, as APs are arranged in a hexagonal grid.

\section{Precoder design}
\subsection{Proposed precoder}
We propose the joint sensing precoder at voxel index $q$, $\mathbf{w}_{q}$, and $\boldsymbol{\rho}_q$ by maximizing (\ref{eqn:SCNR_sensing}), which is equivalent to
\begin{subequations}
    \label{eqn:opt_sensing2}
    \begin{align}
    &\max_{\mathbf{w}_{q},\boldsymbol{\rho}_q } \; \gamma_{\text{sen},q}' = \frac{\mathbf{w}^H_{q}\mathbf{A}\mathbf{w}_{q}}{\xi(\boldsymbol{\rho}_{q})} \\
    &\text{s.t.} \; \norm{\mathbf{w}_{j,q}}^{2} + \rho_{j,q} \leq P_{\text{max}}, \ \forall j,\\
    &{\mathbf{w}^H_{j,q} \mathbf{f}^\ast_{q} }  = 0 ,\ \forall j,\\
    &\gamma_{\text{ue},j,q} \geq \gamma_{\text{req}}, \ \forall j, \\
    &\mathbf{w}^H_{j,q}\textbf{h}_{0,j,\text{d}}^\ast = 0 , \ \forall j, 
    \end{align}
\end{subequations}
where $\mathbf{A} $$\in$$ \mathbb{C}^{JM \times JM}$ is $\mathbf{A}$$=$$ \mathbf{H}_{q}^H\mathbf{v}_q\mathbf{v}_q^H\mathbf{H}_{q}$, $\xi(\boldsymbol{\rho}_q)$$=$$\beta_{\text{g},q}\sum_{j=1}^{J}\rho_{j,q} + \sigma_n^2$, $P_{\text{max}}$ is the maximum transmitting power per $\text{AP}$, $\gamma_{\text{req}}$ is the minimum required user $\text{SINR}$ for the \Gls{ssb} signal at the AP edge, $\boldsymbol{\rho}_q$$=$$[\rho_{1,q},\cdots,\rho_{J,q}]^T$ (\ref{eqn:opt_sensing2}b) is the total transmitted power constraint at each AP, (\ref{eqn:opt_sensing2}c) is an orthogonality constraint between the sensing and the \Gls{ssb} precoders, (\ref{eqn:opt_sensing2}d) is the user SINR constraint, and (\ref{eqn:opt_sensing2}e) is the constraint to mask the direct link from the sensing signal at $\text{AP}_\text{rx}$. Here we refer to the case with direct link suppression, with (\ref{eqn:opt_sensing2}e), as DL-masked, and the case without (\ref{eqn:opt_sensing2}e) as non-DL-masked. Note that  $\mathbf{A} $ is positive semidefinite ($\mathbf{A} $$\succeq$$ 0$), and $\xi(\boldsymbol{\rho}_{q})$ is a linear function of $\boldsymbol{\rho}_{q}$. The problem in (\ref{eqn:opt_sensing2}) is  non-convex, thus we introduce a slack variable $t$ and relax problem (\ref{eqn:opt_sensing2}) using semidefinite relaxation, and rewrite (\ref{eqn:opt_sensing2}) as
\begin{subequations}
    \label{eqn:opt_sdp}
    \begin{align}
    &\max_{\mathbf{W}_q,\boldsymbol{\rho}_q} \; t  \\
    &\text{s.t.} \Tr{ \{ \mathbf{AW}_q \} } \geq t \cdot \xi(\boldsymbol{\rho}_q) \\
    &\Tr{\{[\cdot \; \mathbf{I} \; \cdot]_{j}\mathbf{W}_q [\cdot \; \mathbf{I} \; \cdot]_{j}^T\}} + \rho_{j,q} \leq P_{\text{max}}, \ \forall j,\\
    &\mathbf{f}^T_{q} [\cdot \; \mathbf{I} \; \cdot]_{j}\mathbf{W}_q [\cdot \; \mathbf{I} \; \cdot]_{j}^T \mathbf{f}_{q}^\ast = 0 ,\ \forall j,\\
    &\gamma_{\text{ue},j,q} \geq \gamma_{\text{req}}, \ \forall j, \\
    &\mathbf{h}^T_{0,j,\text{d}} [\cdot \; \mathbf{I} \; \cdot]_{j}\mathbf{W}_q [\cdot \; \mathbf{I} \; \cdot]_{j}^T \mathbf{h}_{0,j,\text{d}}^\ast = 0 ,\ \forall j,
    \end{align}
\end{subequations}
where $\mathbf{W}_q $$\in$$ \mathbb{C}^{JM \times JM}$ is $\mathbf{W}_q$$=$$\mathbf{w}_{q}\mathbf{w}_{q}^H$, $\Tr{ \{ \mathbf{AW}_q \} }$$=$$\mathbf{w}^H_{q}\mathbf{A}\mathbf{w}_{q}$, $[\cdot \; \mathbf{I} \; \cdot]_{j} $$\in$$ \mathbb{C}^{M \times JM}$ is a block-row matrix where $\mathbf{I}$ appears in the $j$th position, and all other blocks are zero matrices, i.e., $[\cdot \; \mathbf{I} \; \cdot]_{3}$$=$$[\mathbf{0} \; \mathbf{0} \; \mathbf{I}]$. The optimization problem in (\ref{eqn:opt_sdp}) can be solved using the convex optimization toolbox (CVX) for $\mathbf{W}_q,\boldsymbol{\rho}_q$ and bisection method for solving $t$. Let  $\mathbf{W}_{\text{sol},q}$ be the solution from the optimizer in (\ref{eqn:opt_sdp}). The eigen-decomposition of $\mathbf{W}_{\text{sol},q}$ is 
\begin{equation}
    \label{eqn:eigendecom}
    \mathbf{W}_{\text{sol},q} = \mathbf{Z}_q \mathbf{\Lambda}_q \mathbf{Z}^{H}_q,
\end{equation}
where $\mathbf{\Lambda}_q$ is a diagonal matrix the containing eigenvalues and the columns in $\mathbf{Z}_q$ are the eigenvectors. The optimized precoder of voxel index $q$ at $\text{AP}_j$, $\mathbf{w}_{\text{sol},q,j} $$\in$$ \mathbb{C}^{M}$, is proportional to the dominant eigenvector in (\ref{eqn:eigendecom}) as
 \begin{equation}
     \label{eqn:precoder_j}
     \mathbf{w}_{\text{sol},q,j} = \sqrt{P_{\text{max}} -\rho_{\text{sol},q,j}} \mathbf{z}_{\text{dom},q,j},
 \end{equation}
 where $\mathbf{z}_{\text{dom},q,j} $$=$$ [\cdot \; \mathbf{I} \; \cdot]_{j}\mathbf{z}_{\text{dom},q}$, $\mathbf{z}_{\text{dom},q} $$\in$$ \mathbb{C}^{JM}$ is the dominant eigenvector of $\mathbf{W}_{\text{sol},q}$ in (\ref{eqn:eigendecom}), and $\rho_{\text{sol},q,j}$ is the optimized power of SSB signal from (\ref{eqn:opt_sdp}). Note that the optimized precoder is suboptimal because the optimization problem has been relaxed.

 \subsection{Non-coordinated precoder}
The non-coordinated precoder is formulated independently at each $\text{AP}$ and is used as a benchmark to evaluate the performance of the proposed precoder. The non-coordinated precoder at $\text{AP}_j$ for the voxel index $q$ is selected as a steering vector that projects onto the orthogonal complement of the column space of $\mathbf{U}_{j,q}$ as $\mathbf{w}_{j,q} = e^{j2\pi \varphi_{j,q}} \sqrt{\Tilde{\rho}_{j,q}} \Tilde{\mathbf{w}}_{j,q}/\norm{\Tilde{\mathbf{w}}_{j,q}}$, where
\begin{equation}
    \label{eqn:uncorrdinated_prec}
    \Tilde{\mathbf{w}}_{j,q} = \big(\mathbf{I}-\mathbf{U}_{j,q}(\mathbf{U}_{j,q}^H\mathbf{U}_{j,q})^{-1}\mathbf{U}_{j,q}^H \big) {\mathbf{h}}_{j,q},
\end{equation}
$\Tilde{\rho}_{j,q}$ represents the sensing power, which is kept identical in the proposed precoder to ensure a fair comparison, $\mathbf{U}_{j,q}$$=$$[\Tilde{\mathbf{h}}_{0,j,\text{d}},\mathbf{f}^\ast_q]$, $\Tilde{\mathbf{h}}_{0,j,\text{d}}$$=$$\mathbf{h}_{0,j,\text{d}}/\sqrt{M}$ is the normalized direct link departure channel, and $\mathbf{f}^\ast_q$ is the normalized SSB precoder. The non-coordination is highlighted in $\varphi_{j,q}$, which represents the phase difference associated with a wavelength factor from the reference $\text{AP}_1$ and is denoted as 
\begin{equation}
    \label{eqn:phase_diff}
    \varphi_{j,q}=\frac{{l_{j,q}-l_{1,q}}}{\lambda},
\end{equation}
where $\lambda$ is the wavelength and ${l_{j,q}-l_{1,q}}$ is the difference in the distance from $\text{AP}_j$ and $\text{AP}_1$ to voxel $q$. The columns of $\mathbf{U}_{j,q}$ are approximately orthonormal because the directions of the SSB beams and the direct link are sufficiently separated, and a large number of antennas is used; hence, (\ref{eqn:uncorrdinated_prec}) reduces to 
\begin{equation}
    \label{eqn:uncorrdinated_prec_red}
    \Tilde{\mathbf{w}}_{j,q} = (\mathbf{I}-\mathbf{U}_{j,q} \mathbf{U}_{j,q}^H)\mathbf{h}_{j,q}.
\end{equation}

\section{Drone detection}
In this section, we present the detector design for each voxel. The detection process is formulated as a hypothesis testing problem, where the decision is based on comparing the observed signal against predefined hypotheses. Under hypothesis $\mathcal{H}_0$ (drone absent), the target paths are absent ($\eta=0$) while retaining clutter, and noise. Under hypothesis $\mathcal{H}_1$ (drone present), the target paths exist $(\eta=1)$ and include clutter, and noise. The signal models for both hypotheses at the voxel $q$ are
\begin{equation}
    \label{eqn:hypotheses}
\begin{split}
    &\mathcal{H}_0: \mathbf{y}_q= \mathbf{x}_{\text{c},q} + \mathbf{n}_q, \\
    &\mathcal{H}_1: \mathbf{y}_q=\mathbf{x}_{\text{s},q}+\mathbf{x}_{\text{c},q}+\mathbf{n}_q,
\end{split}
\end{equation}
where $\mathbf{y}_q$ is the received signal at $\text{AP}_\text{rx}$, $\mathbf{x}_{\text{s},q}$$\sim$$ \mathcal{CN}(0,\sigma_\text{rcs}^2 \mathbf{\Phi}_q)$ is the sensing component (the first term in (\ref{eqn:rx_apr})), $\sigma_\text{rcs}^2 \mathbf{\Phi}_q$ is the covariance matrix of the sensing component, $\mathbf{\Phi}_q\in \mathbb{C}^{M\times M}$, $\mathbf{\Phi}_q$$=$$\sum_j^J\mathbf{H}_{j,q}\mathbf{w}_{j,q}\mathbf{w}_{j,q}^H\mathbf{H}^H_{j,q}$, $\mathbf{x}_{\text{c},q}$$\sim$$ \mathcal{CN}(0,\rho_q\beta_{\text{g},q} \mathbf{I})$ is the clutter component (second term in (\ref{eqn:rx_apr})), $\mathbf{n}_{q}$$\sim$$ \mathcal{CN}(0,\sigma_\text{n}^2 \mathbf{I})$ is noise, and the direct link component is disregarded. The likelihood function for both hypotheses are
\begin{equation}
    \label{eqn:likelihood}
    \begin{split}
        p(\mathbf{y}_q|\mathcal{H}_0) &= \mathcal{CN}(0,\sigma^2\mathbf{I}) \\
        &= \frac{1}{\pi^M \sigma^{2M}} 
\exp\!\Big( -\sigma^{-2} \mathbf{y}_q^H\mathbf{y}_q \Big), \\
        p(\mathbf{y}_q|\mathcal{H}_1) &= \mathcal{CN}(0,\sigma^2\mathbf{I}+\sigma_\text{rcs}^2\mathbf{\Phi}_q) = \mathcal{CN}(0,\mathbf{C}_q)\\ 
        &=\frac{1}{\pi^M \det(\mathbf{C}_q)}
         \exp\!\Big(-\mathbf{y}_q^H\mathbf{C}_q^{-1} \mathbf{y}_q\Big),
    \end{split}
\end{equation}
where $\sigma^2=\sigma^2_\text{n} +\rho_q\beta_{\text{g},q}$, $\rho_q=\sum_j \rho_{j,q}$, and $\mathbf{C}_q=\sigma^2\mathbf{I}+\sigma_\text{rcs}^2\mathbf{\Phi}_q$. The log-likelihood ratio, $\operatorname{LLR}(\mathbf{y}_q)$,   is
\begin{equation}
\mathbb{\label{eqn:LLR}}
    \begin{split}
        \operatorname{LLR}(\mathbf{y}_q) &= \ln \Bigg(\frac{p(\mathbf{y}_q|\mathcal{H}_1)}{p(\mathbf{y}_q|\mathcal{H}_0)}\Bigg)\\
        &=\ln\Bigg(\frac{\sigma^{2M}}{\det(\mathbf{C}_q)}\Bigg)+ \mathbf{y}_q^H(\sigma^{-2}\mathbf{I}-\mathbf{C}_q^{-1})\mathbf{y}_q.
    \end{split}
\end{equation}
The test statistic is
\begin{equation}
\begin{split}
\label{eqn:Tstatis}
T(\mathbf{y}_q)=\mathbf{y}_q^H(\sigma^{-2}\mathbf{I}-\mathbf{C}_q^{-1})\mathbf{y}_q \underset{\mathcal{H}_0}{\overset{\mathcal{H}_1}{\gtrless}}
 \gamma_\text{th},\\
\end{split}
\end{equation}
where the constant term is omitted. The optimal test can be derived as in \cite[Chap.~5]{Kay1998} as
\begin{equation}
    \begin{split}
T(\mathbf{y}_q)=\mathbf{y}_q^H\mathbf{\Phi}_q(\sigma^{-2}\mathbf{I}+\sigma^2_\text{rcs}\mathbf{\Phi}_q)^{-1}\mathbf{y}_q \underset{\mathcal{H}_0}{\overset{\mathcal{H}_1}{\gtrless}}
 \gamma_\text{th}',
    \end{split}
\end{equation}
where $\gamma_\text{th}'=\frac{\sigma^2}{\sigma^2_\text{rcs}}\gamma_\text{th}$.
\section{Simulation results}
\label{sec:sim}

\begin{figure}[htbp] 
    \centering
        \resizebox{0.8\linewidth}{!}{
%
%
\definecolor{mycolor1}{rgb}{1.00000,0.00000,1.00000}%
\definecolor{colorb}{RGB}{33,113,181}   
\definecolor{colorm}{RGB}{231,41,138}   
\definecolor{colorg}{RGB}{52,160,44}    
\definecolor{colorp}{RGB}{117,112,179}  
\definecolor{coloro}{RGB}{255,127,0}    
\begin{tikzpicture}

\begin{axis}[%
width=4.521in,
height=3.528in,
at={(0.758in,0.519in)},
scale only axis,
xmin=-5,
xmax=30,
xlabel style={font=\LARGE\color{white!15!black}},
xlabel={$P_\mathrm{max}$ (dBm)},
ymode=log,
ymin=0.001,
ymax=100000,
yminorticks=true,
ylabel style={font=\LARGE\color{white!15!black}},
ylabel={$\gamma_\mathrm{sen}$},
axis background/.style={fill=white},
xmajorgrids,
ymajorgrids,
yminorgrids,
grid style={line width=.1pt, draw=gray!30}, 
legend style={at={(0.03,0.97)}, anchor=north west, legend cell align=left, align=left, draw=white!15!black}
]
\addplot [color=coloro, mark=square, mark repeat=10, line width=3.0pt]
  table[row sep=crcr]{%
-1.96311106027476	0\\
-1.86311106027476	0.00866143093044424\\
-1.76311106027476	0.0172613069369959\\
-1.66311106027476	0.0261065752031199\\
-1.56311106027476	0.0349006442196082\\
-1.46311106027476	0.0468028697577433\\
-1.36311106027476	0.0545137196114917\\
-1.26311106027476	0.0634792733895938\\
-1.16311106027476	0.0734333803435637\\
-1	0.0958949652728902\\
-0.5	0.151568140475566\\
0	0.212807245669618\\
0.5	0.28077120106382\\
1	0.374274203903854\\
1.5	0.442153980193962\\
2	0.540721164590967\\
2.5	0.669596797202817\\
3	0.793234351595206\\
3.5	0.925098916668275\\
4	1.10431307554857\\
4.5	1.29071379112844\\
5	1.47712704483973\\
5.5	1.67062141233642\\
6	1.94103610295176\\
6.5	2.26208782960263\\
7	2.56107360778838\\
7.5	2.89364115597247\\
8	3.24502119475766\\
8.5	3.7755704907187\\
9	4.14267356338674\\
9.5	4.76033171826409\\
10	5.39995843010106\\
10.5	6.08096688132092\\
11	6.68771457469358\\
11.5	7.66166608917349\\
12	8.73162951303515\\
12.5	9.99317789938568\\
13	11.109433987273\\
13.5	12.1397030631991\\
14	14.080386620125\\
14.5	15.8812409003663\\
15	18.2493173221877\\
15.5	20.2856195889891\\
16	23.1132770150628\\
16.5	25.548412001049\\
17	29.1121873155229\\
17.5	31.9519979681301\\
18	36.0993037960781\\
18.5	41.5027048686475\\
19	45.9678538258009\\
19.5	51.6659881137963\\
20	57.3876149074987\\
20.5	65.5264595664369\\
21	70.6747242292572\\
21.5	81.9992381784867\\
22	89.6088068840719\\
22.5	104.854298408458\\
23	117.553734881267\\
23.5	129.282423726117\\
24	147.363010885855\\
24.5	166.238009013861\\
25	179.785940395184\\
25.5	206.368479074556\\
26	227.955357702413\\
26.5	257.983105216909\\
27	292.975915760872\\
27.5	319.17715547968\\
28	378.532027030255\\
28.5	411.574217382488\\
29	449.287659880912\\
29.5	514.10775911866\\
30	583.706847860442\\
};
\addlegendentry{$\gamma_{\mathrm{req}} = 2 $ dB, Proposed precoder}

\addplot [color=colorb, mark=triangle, mark repeat=10, line width=3.0pt]
  table[row sep=crcr]{%
18.462825648618	6.74605207228138e-15\\
18.562825648618	0.769071693727543\\
18.662825648618	1.5829200552782\\
18.762825648618	2.38738328230075\\
18.862825648618	3.12095586690654\\
19	4.38129615110802\\
19.5	8.81387036328001\\
20	14.3552939326102\\
20.5	19.7021864668728\\
21	27.0445627194946\\
21.5	33.1081750499238\\
22	43.5775284835622\\
22.5	52.7193621096574\\
23	61.118601785135\\
23.5	73.1303030224474\\
24	85.382959564027\\
24.5	98.2416280341567\\
25	118.623737865772\\
25.5	134.451231409676\\
26	153.471369407288\\
26.5	179.517385281881\\
27	201.731599922421\\
27.5	235.674088464556\\
28	268.849202270081\\
28.5	303.461397129879\\
29	343.395161452254\\
29.5	394.043307242614\\
30	438.675664610715\\
};
\addlegendentry{$\gamma_{\mathrm{req}} = 3 $ dB, Proposed precoder}

\addplot [color=colorg,mark=diamond,mark repeat=10, line width=3.0pt]
  table[row sep=crcr]{%
-1.96311106027476	0\\
-1.86311106027476	7.23715641337101e-05\\
-1.76311106027476	0.000145247478232023\\
-1.66311106027476	0.000232827288589636\\
-1.56311106027476	0.000298026755778238\\
-1.46311106027476	0.000388996164872808\\
-1.36311106027476	0.000461848379098159\\
-1.26311106027476	0.000536527108275455\\
-1.16311106027476	0.000647399960606655\\
-1	0.000788728422829915\\
-0.5	0.00126720951917954\\
0	0.00186119546057072\\
0.5	0.00241283995127811\\
1	0.00305591273872313\\
1.5	0.00390897102388301\\
2	0.00469050453194911\\
2.5	0.00570890375817924\\
3	0.00676785936969595\\
3.5	0.00789592635104187\\
4	0.0089269428530802\\
4.5	0.010758278242451\\
5	0.0123903967456778\\
5.5	0.0142549238647054\\
6	0.0165036169303614\\
6.5	0.0188907070475955\\
7	0.0216339131565923\\
7.5	0.0247427563036805\\
8	0.0274907680168664\\
8.5	0.0320473433338195\\
9	0.0346967744295518\\
9.5	0.0401132877461856\\
10	0.0445618236466841\\
10.5	0.0530289703795249\\
11	0.0604872452493749\\
11.5	0.0668029691981191\\
12	0.0735181171031674\\
12.5	0.0827416828393239\\
13	0.093009260090524\\
13.5	0.104125686922457\\
14	0.116133929971902\\
14.5	0.134790875445475\\
15	0.151457073038461\\
15.5	0.16846470982357\\
16	0.187763374706229\\
16.5	0.217574749270205\\
17	0.240386275621782\\
17.5	0.257692124519621\\
18	0.308442349972777\\
18.5	0.333604392836994\\
19	0.387530449326697\\
19.5	0.417361518700616\\
20	0.480804648486538\\
20.5	0.527849703083292\\
21	0.598975876620043\\
21.5	0.678783012470393\\
22	0.718524619624065\\
22.5	0.84196185388415\\
23	0.921655158939522\\
23.5	0.994336046674034\\
24	1.14081838427267\\
24.5	1.24754750692957\\
25	1.37422843905249\\
25.5	1.51131808554383\\
26	1.66610360536358\\
26.5	1.86524902797774\\
27	2.12338144831421\\
27.5	2.29670519781289\\
28	2.57157599534822\\
28.5	2.73897845999424\\
29	2.98212080692192\\
29.5	3.27913978525832\\
30	3.57267022539519\\
};
\addlegendentry{$\gamma_{\mathrm{req}} = 2 $ dB, Non-coordinated precoder}

\addplot [color=black,mark=o, mark repeat=10, line width=3.0pt]
  table[row sep=crcr]{%
18.462825648618	5.77917346274081e-17\\
18.562825648618	0.00661760952310464\\
18.662825648618	0.0138165305290616\\
18.762825648618	0.0198515569521286\\
18.862825648618	0.0269395616725491\\
19	0.0363911446186386\\
19.5	0.0766250175526124\\
20	0.116628174635875\\
20.5	0.166627480641063\\
21	0.223753623593759\\
21.5	0.273464040067405\\
22	0.345382756833876\\
22.5	0.42885181844534\\
23	0.491622250152699\\
23.5	0.593147226450942\\
24	0.698589293762156\\
24.5	0.793744448386231\\
25	0.928832499280201\\
25.5	1.07172324532238\\
26	1.19942523832467\\
26.5	1.39578363742607\\
27	1.5593847823852\\
27.5	1.75783814191393\\
28	1.91409037087915\\
28.5	2.17428727239343\\
29	2.33800245160047\\
29.5	2.61908186900513\\
30	2.88357882275019\\
};
\addlegendentry{$\gamma_{\mathrm{req}} = 3 $ dB, Non-coordinated precoder}

\end{axis}

\begin{axis}[%
width=5.833in,
height=4.375in,
at={(0in,0in)},
scale only axis,
xmin=0,
xmax=1,
ymin=0,
ymax=1,
axis line style={draw=none},
ticks=none,
axis x line*=bottom,
axis y line*=left
]
\end{axis}
\end{tikzpicture}%
        }
        \caption{Sensing SINR.}
    \label{fig:result1}
\end{figure}
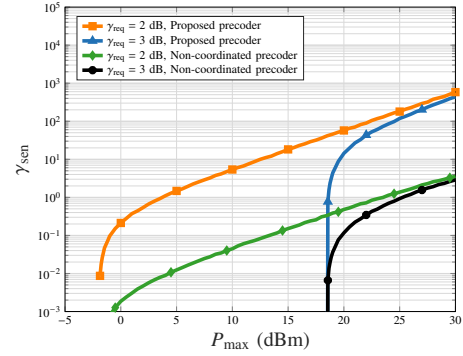

\begin{table}
    \caption{Baseline parameters.}
    \begin{center}
        \begin{tabular}{c|c|c|c   }
            \hline
            \textbf{Parameter}  & \textbf{Value} & \textbf{Parameter}  & \textbf{Value} \\
            \hline
            $M_\text{V} $$,$$ M_\text{H}$ & $12$ &      $\sigma_\text{RCS}$ & $-10$ dBsm \\
            \hline
            $f_c$ & $15$ GHz & $z$ & $10$ m  \\
            \hline
            $\beta_{\text{g},q}$ & $-90$dB  & $\sigma^2_\text{n}$ & -60 dBm
            \\ 
            \hline
            $P_{\text{max}}$ & $30$ dBm &  $d$ & $2$ $m$   \\
            \hline
            $\gamma_{\text{req}}$ & $3$ dB & Volume  & 6x2x2 $m^3$ \\
            \hline
             r & $250$ m \\
            \hline
            
        \end{tabular}
        \label{tab:params}
    \end{center}
    \label{table:simparams}
\end{table}
We first evaluate the sensing performance of the proposed precoder against the non-coordinated precoder by using the averaged sensing SINR, $\gamma_{\text{sen}}$, over all voxels within the volume. The baseline parameters and values for all simulations are shown in Table \ref{tab:params}, unless indicated otherwise. In Fig. \ref{fig:result1}, $\gamma_{\text{sen}}$ begins to be observed at $P_{\text{max}}$$=$$-2,\ \text{and } 18.5$ dBm corresponding to the required user SINR $\gamma_{\text{req}}$$=$$2,\ \text{and}\ 3$ dB, respectively. The proposed precoder achieves an improvement of $\gamma_\text{sen}$ approximately $20$ dB over the non-coordinated precoder. In addition, $\gamma_\text{sen}$ increases linearly with $P_{\text{max}}$ once the user requirement constraint has been satisfied.

\begin{figure}[htbp] 
    \centering
        \includegraphics[width=2.3in]{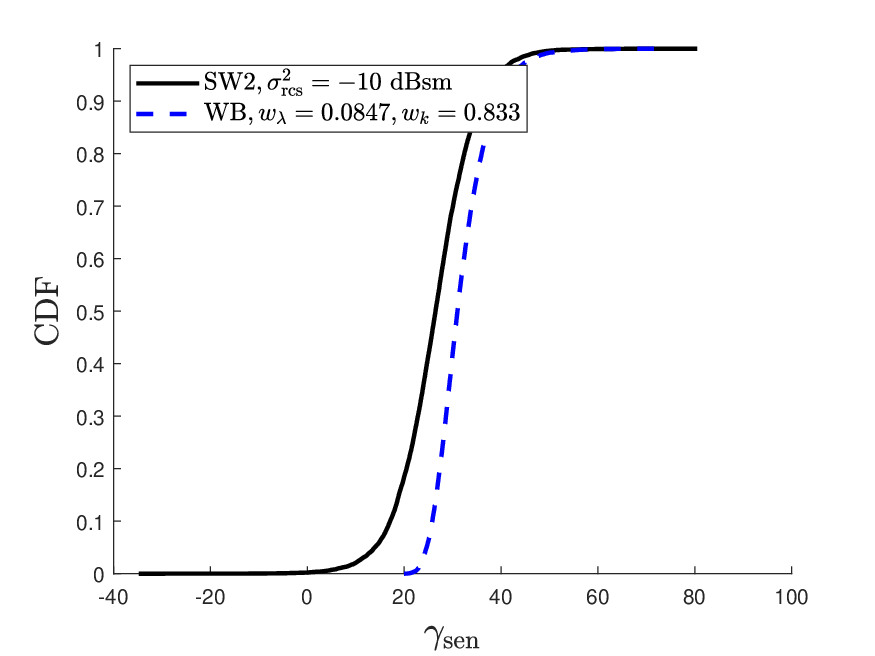}
        \caption{Empirical CDF of $\gamma_\text{sen}$.}
    \label{fig:result1_1}
\end{figure}

A study in \cite{10106375} investigates the statistical characteristics of the RCS of a fixed-wing drone during flight. We apply the proposed RCS model in \cite{10106375}, Weibull distribution (\emph{WB}) with given scale and shape parameters, and compare the cumulative distribution function (CDF) of the $\gamma_{\text{sen}}$ with \emph{SW2} with comparative variance ($\sigma_\text{RCS}$$=$$-10$ dBsm), as shown in Fig.~\ref{fig:result1_1}. The sensing SINR of \emph{WB} RCS yields a slightly steeper CDF compared to that of \emph{SW2}. This highlights that applying \emph{SW2} with an RCS variance of $-10$ dBsm yields lower performance compared to the given \emph{WB}. Therefore, evaluating the performance using \emph{SW2} guarantees an outcome that is conservative in practice while still relying on simplified signal modeling.

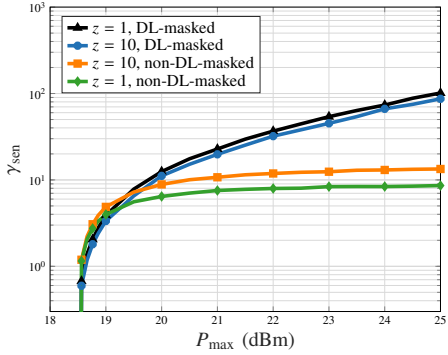
\begin{figure}[htbp] 
    \centering
        \resizebox{0.8\linewidth}{!}{
%
%
\definecolor{mycolor1}{rgb}{1.00000,0.00000,1.00000}%
\definecolor{colorb}{RGB}{33,113,181}   
\definecolor{colorm}{RGB}{231,41,138}   
\definecolor{colorg}{RGB}{52,160,44}    
\definecolor{colorp}{RGB}{117,112,179}  
\definecolor{coloro}{RGB}{255,127,0}    
\begin{tikzpicture}

\begin{axis}[%
width=4.521in,
height=3.528in,
at={(0.758in,0.519in)},
scale only axis,
xmin=18,
xmax=25,
xlabel style={font=\LARGE\color{white!15!black}},
xlabel={$P_\mathrm{max}$ (dBm)},
ymode=log,
ymin=0.3,
ymax=1000,
yminorticks=true,
ylabel style={font=\LARGE\color{white!15!black}},
ylabel={$\gamma_\mathrm{sen}$},
axis background/.style={fill=white},
xmajorgrids,
ymajorgrids,
yminorgrids,
grid style={line width=.1pt, draw=gray!30}, 
legend style={font=\Large, at={(0.03,0.97)}, anchor=north west, legend cell align=left, align=left, draw=white!15!black}
]
\addplot [color=black,mark=triangle,mark repeat=2, line width=3.0pt]
  table[row sep=crcr]{%
18.462825648618	5.86798310188578e-15\\
18.562825648618	0.673885288327054\\
18.662825648618	1.39647585134496\\
18.762825648618	2.04912015029335\\
18.862825648618	2.78174697056724\\
19	3.90090149617948\\
19.5	7.82013566862104\\
20	12.4882967263125\\
20.5	17.5590104290622\\
21	22.7779235491125\\
21.5	29.6316675478137\\
22	36.6970170413347\\
22.5	44.5613709410422\\
23	53.912002275231\\
23.5	63.6055040037462\\
24	73.529002906996\\
24.5	88.109609879633\\
25	101.330257657468\\
25.5	116.309491619909\\
26	134.391514658583\\
26.5	152.524959523493\\
27	178.100901935476\\
27.5	199.731167774216\\
28	229.338410920043\\
28.5	262.530616306073\\
29	301.803767273333\\
29.5	343.960013543254\\
30	379.503271577211\\
};
\addlegendentry{$z=1, $ DL-masked}

\addplot [color=colorb, mark=o,mark repeat=2, line width=3.0pt]
  table[row sep=crcr]{%
18.462825648618	5.29013837971596e-15\\
18.562825648618	0.596049136142919\\
18.662825648618	1.18037285191368\\
18.762825648618	1.80176718357674\\
18.862825648618	2.40780739861899\\
19	3.3485461241709\\
19.5	6.52467464057624\\
20	11.140382253431\\
20.5	15.1686953680388\\
21	19.7789002691798\\
21.5	25.2710319681626\\
22	32.0949563676221\\
22.5	38.0345406845864\\
23	45.1519052491774\\
23.5	53.9951680211775\\
24	66.4656043638022\\
24.5	74.8985430639052\\
25	86.8232698596321\\
25.5	104.954851001922\\
26	114.410282128422\\
26.5	129.314373462149\\
27	157.360778971563\\
27.5	171.091953772573\\
28	195.423441427985\\
28.5	228.270304149027\\
29	260.810037760085\\
29.5	298.410222785272\\
30	335.526469476338\\
};
\addlegendentry{$z=10, $ DL-masked}

\addplot [color=coloro, mark=square,mark repeat=2, line width=3.0pt]
  table[row sep=crcr]{%
18.4628256486179	0\\
18.5628256486179	1.19156255483527\\
18.6628256486179	2.2137176005621\\
18.7628256486179	3.06814319545708\\
18.8628256486179	3.8911593751163\\
19	4.87424796411789\\
19.5	7.2284468432539\\
20	8.87188983451716\\
20.5	10.0467780235254\\
21	10.7293302435175\\
21.5	11.4636063519465\\
22	11.8734245260755\\
22.5	12.2566355721045\\
23	12.4378626010358\\
23.5	12.9369588487605\\
24	13.0356230135363\\
24.5	13.2681584652719\\
25	13.3809653086046\\
25.5	13.5489255695154\\
26	13.6634273973244\\
26.5	13.8412492013341\\
27	13.8220991274657\\
27.5	13.5993787238884\\
28	13.7919139648177\\
28.5	14.0814681500579\\
29	13.8555764376688\\
29.5	14.1610902546075\\
30	14.0461380710259\\
};
\addlegendentry{$z=10, $ non-DL-masked}

\addplot [color=colorg,mark=diamond,mark repeat=2, line width=3.0pt]
  table[row sep=crcr]{%
18.462825648618	1.09993478122233e-14\\
18.562825648618	1.1526544496163\\
18.662825648618	2.06026907100156\\
18.762825648618	2.74161886174252\\
18.862825648618	3.33217242606382\\
19	3.98291864597383\\
19.5	5.5807849674315\\
20	6.43535448736411\\
20.5	7.03654115535384\\
21	7.54168448766808\\
21.5	7.78246405918244\\
22	7.96150740780398\\
22.5	8.02242777918687\\
23	8.36297129271611\\
23.5	8.39112762224355\\
24	8.38009095666799\\
24.5	8.46830035978484\\
25	8.60373350977659\\
25.5	8.74758874452278\\
26	8.64882517837635\\
26.5	8.61316730080951\\
27	8.62812618782926\\
27.5	8.76781008856815\\
28	8.81152068553698\\
28.5	8.82806270861248\\
29	8.9175973215309\\
29.5	8.75833894000039\\
30	8.90649489215125\\
};
\addlegendentry{$z=1, $ non-DL-masked}

\end{axis}

\begin{axis}[%
width=5.833in,
height=4.375in,
at={(0in,0in)},
scale only axis,
xmin=0,
xmax=1,
ymin=0,
ymax=1,
axis line style={draw=none},
ticks=none,
axis x line*=bottom,
axis y line*=left
]
\end{axis}
\end{tikzpicture}%
        }
        \caption{$\gamma_\text{sen}$ at different altitudes (z levels).}
    \label{fig:result2}
\end{figure}

Fig. \ref{fig:result2} shows the evaluation of constraint (\ref{eqn:opt_sensing2}.e) at different volumetric altitudes $(z)$. The result shows $\gamma_\text{sen}$ saturation in the non-DL-masked cases for both $z$ levels as $P_\text{max}$ increases, while in the DL-masked case, $\gamma_\text{sen}$ increases linearly with $P_\text{max}$. Moreover, the separation in $\gamma_\text{sen}$ between the levels of $1$ m and $10$ m $z$ is relatively small in the DL-masked case compared to the non-DL-masked case. This highlights the minimal effect of varying altitudes in the DL-masked case.

\begin{figure}[htbp] 
    \centering
        \resizebox{0.8\linewidth}{!}{
        \input{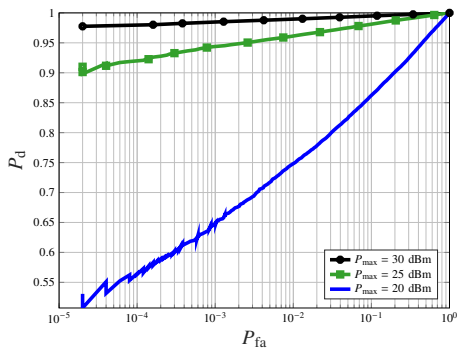}
        }
        \caption{Detection performance on a single voxel.}
    \label{fig:result3}
\end{figure}

Fig.~\ref{fig:result3} illustrates the detection performance of the proposed precoder for different values of $P_\text{max}$ on a single voxel given $q=1$. The receiver operating characteristic (ROC) indicates relatively good detection performance for $P_\text{max}=30$ dBm and $25$ dBm, while a noticeable degradation occurs for $P_\text{max}=20$, as most of the power is allocated to the SSB signal. For $P_\text{max}=25$ dBm, the probability of detection $(P_\text{d})$ is about $0.9$ with the probability of false alarm $P_\text{fa}=2\cdot10^{-5}$, while for $P_\text{max}=30$ dBm, $P_\text{d}$ is about $0.97$ with the same $P_\text{fa}$. This highlights a significant improvement in $P_\text{d}$ from $P_\text{max}=20$ dBm to $P_\text{max}=25$ dBm, whereas the improvement from $P_\text{max}=25$ dBm to $P_\text{max}=30$ dBm is marginal, given the same $P_\text{fa}=2\cdot10^{-5}$. This saturation occurs because, at high SINR, the distributions under $\mathcal{H}_0$ and $\mathcal{H}_1$ are well separated.

\section{Conclusion}
This paper proposes a novel \Gls{isac} scheme for low-altitude drone detection within the current 5G system. The scheme leverages reference-type signal slots, specifically the \Gls{ssb} signal. During beam sweeping of the SSB, the sensing signal is transmitted toward a predefined grid of voxels that defines the search volume. The precoder is designed to minimize the impact of sensing interference and to satisfy the UE SINR. The results demonstrate that the proposed precoder outperforms the non-coordinated precoder and is minimally affected by variations in drone altitude.

\section*{Acknowledgment}
This work was supported in part by ELLIIT and in part by the WASP-funded project ``ALERT''.

\bibliographystyle{IEEEtran}
\bibliography{reference}
\end{document}